\documentclass[twocolumn]{article}

\usepackage[dvips]{graphicx}
\usepackage{amsmath,amssymb,amsfonts,verbatim}
\usepackage{calc}
\usepackage[T1]{fontenc}
\usepackage{color}

\begin{document}

\title{Local detection of X-ray spectroscopies with an in-situ AFM}

\author{\normalsize M. S. Rodrigues$^1$, O. Dhez$^1$, S. Le Denmat$^1$, J. Chevrier$^2$, R. Felici$^1$ and F. Comin$^1$ \\ \small $^1$ European Synchrotron Radiation Facility, Grenoble, France \\ \small $^2$ Institut Néel, CNRS/UJF Grenoble, France\\
\\
\small{Email: mario.rodrigues@esrf.fr} \\
\\
\begin{minipage}[h]{14 cm}
\small
The in situ combination of Scanning Probe Microscopies (SPM) with X-ray microbeams adds a variety of new possibilities to the panoply of synchrotron radiation techniques.
In this paper we describe an optics-free AFM/STM that can be directly installed on synchrotron radiation end stations for such combined experiments. The instrument can be used just for AFM imaging of the investigated sample or can be used for detection of photoemitted electrons with a sharp STM-like tip, thus leading to the local measure of the X-ray absorption signal. Alternatively one can can measure the flux of photon impinging on the sharpest part of the tip to locally map the pattern of beams diffracted from the sample. In this paper we eventually provide some examples of local detection of XAS and diffraction.
\end{minipage}}

\maketitle

\section{Introduction}

SPM techniques are used in many scientific fields ranging from biology to materials sciences.
Nowadays, they are thought to be at the hearth of nanosciences.
They are easy to use, produce high resolution images of the sample and can unveil many local properties of the surface.
Synchrotron Radiation is used instead to probe the atomic and electronic structures of surfaces averaging over the illuminated area.
In the most recent years the use of  micro and nano X-ray beams is steadily increasing and the join exploitation of many SPM or SR techniques appears more and more desirable.

In micro-nano characterization it is often necessary to work on a single object whose size can vary from the micro to the nanoscale. 
It is then covetable to be able to perform SPM and X-ray experiments on the same single object in the same conditions and at the same time for more systematic and comprehensive exploration of the micro-nano world.

So far two paths have been explored.
A first one consists in installing an UHV SPM chamber that can be installed on a dedicated X-ray beam line \cite{Saito, Matsushima, Eguchi}.
The combination of STM and X-rays is aimed to provide chemical contrast in near field microscopies \cite{Gimzewnski}.
Detecting the X-ray induced change in the tunneling probability gives
to the STM atomic resolution in topographic mode, and, according to the authors, chemical contrast down to the tens of nm.

The second path is the one we have chosen at the ESRF in the framework of the X-Tip European Project \cite{Xtip}. We considered that an extensive integration of SPM techniques on SR instrumentation could be done only leaving the beamlines as they are and try to adapt a compact, optics free, AFM/STM on an end station to perform indifferently diffraction or spectroscopy experiments.
Clearly, the aim here is to emphasize versatility, ease of use and the largest possible spectrum of applications. Difficulties are essentially centered on issues of SPM stability, on the procedures for the simultaneous alignment of tip, sample microbeam, and finally lateral resolution.

Even though the AFM tip has been used successfully to locally detect X-rays absorption and diffraction data, the lateral resolution is still rather limited. This limitation is solely due to the present technical difficulty of shielding the tip rendering it blind, except at its apex.
Once this task is accomplished, the expected lateral resolution should be better than 50 nm.

To be mentioned is also the work of Ishii \cite{Ishii} who has taken a different approach using scanning capacitance microscopy to probe localized electrons.

\section{Experimental Set up}

The home built AFM is based on a quartz tuning fork (TF).
Use of tuning forks in SPM design has been the subject of vast literature because they are widely used in SNOM setups \cite{TF_SNOM}. In the meanwhile its use in AFMs has been growing very fast \cite{TF_AFM1, TF_AFM2}.

Our instrument operates very much like a conventional AFM but with a simplified set up: unlike conventional AFMs, where the deflection of a cantilever is detected by a laser beam, AFMs based on TFs have the added simplicity that the motion of the oscillator is detected directly by the conversion of the alternating stress field inside the quartz piezoelectric material into charge. 
Neither a laser nor a photodetector are necessary for the operation of this type of AFMs. This simplifies  this type of AFM and makes it more robust.
However, it may be slower because of the very high quality factor (Q $\approx$ 8000 versus Q $\approx$ 30 for a conventional cantilever). In other words, it shows a very large time constant ($\tau \approx$ 1s) which actually compensates for their rigidity k $\approx$ 20 kN/m (for a conventional cantilever this value is about 40 N/m).
The sensitivity depends on k/Q and this value is roughly the same for a conventional AFM Si cantilever or a TF.

Typical natural frequencies of the TFs after mounting of the tip are around 30kHz.

The high Q factor of the TF has led us to choose PLL (Phase Locked Loop) detection to minimize the time needed to acquire an image.
In few words, the PLL helps the TF to stay close to equilibrium at all times since a feedback loop actively excites the TF at varying frequencies maintaining the phase shift constant at -90 Deg and thus keeping it at its resonance frequency.

The tungsten tip is mounted on the tuning fork by means of a conductive glue to ensure the electrical connection with the TF electrode connected to the preamp.
Assembling a sharp tungsten tip onto an electrically connected TF has become a current procedure in our lab, making SR experiments quite straightforward from this point of view.

\begin{figure}
		\includegraphics[width=1\linewidth]{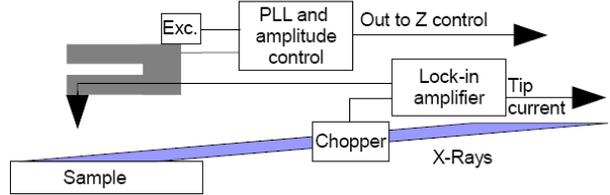}
			\caption{Experimental setup for combination of AFM with X-rays. 					\vspace{-16pt}}
			\label{scheme}
\end{figure}

The sample is illuminated with time modulated monochromatic X-ray light:
a slotted rotating wheel (chopper) is used to modulate the intensity of the beam at a frequency between 1kHz and 5kHz.
The current detected by the tip is demodulated with a lock-in amplifier.
The experimental arrangement is shown in fig. \ref{scheme}.
The X-ray beam used to obtain the results presented below had 10$^{12}$ photons per second in about 10$^{-4}$ bandpass on about 4$\times$4 $\mu$m$^2$.

The sample consisted of Ge islands of about 1$\mu$m width and 500nm height deposited on a Si substrate.
An in-situ AFM image reveals its topography (figure \ref{AFM_EXAFS}a).

\section{Microbeam interaction with the W tip}

The first step of the experiment is the alignment of the beam with the tip of the AFM.

\begin{figure}
	\begin{center}
		\includegraphics[width=1\linewidth]{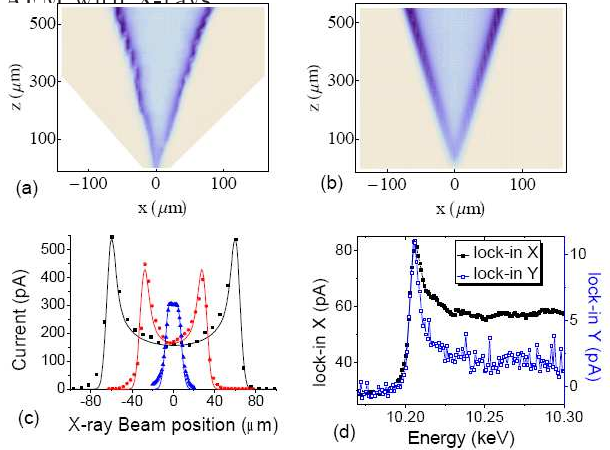}
			\caption{a) 3D plot of the current flowing to the W tip as it is scanned by the beam. b) Simulation of a). c) cross sections of a) and b). d) current from the tip as a function of photon energy across the$L_3$ edge of the W tip. \vspace{-16pt}}
			\label{tip_mesh}
	\end{center}
\end{figure}

The total electron yeld (TEY) due to tip photoemission produces a measurable current which has been systematically recorded as the tip is moved in the XZ plan perpendicular to the beam, thus producing experimental images of the tip as shown in fig. \ref{tip_mesh} (a) with the respective profiles in fig. \ref{tip_mesh} (c).

A characteristic feature of this type of scans is an increase in the current when the X-ray beam is at grazing incidence on the tip. This produces the higher contrast on the tip edges well visible in fig. \ref{tip_mesh}a and \ref{tip_mesh}b.

The relevant parameters to analyze the experimental results of figure \ref{tip_mesh}a are the linear absorption coefficient $\mu_l$ (which is of few microns for the energy used), the inelastic mean free path  for e$^-$ in solids $\lambda$ (which is typically of few nm), and the characteristic energy loss per unit of length i.e., the stopping power S(E) for electrons.

In order to quantify the results, we define the effective escape depth $d_{eff}$ as the depth for which there is a probability $1/2$  that a photon absorbed at a distance smaller than $2/3 d_{eff}$ will result in one electron escaping the tip.
The other e$^-$ go further deep into the tip and do not escape.
The factor $2/3$ corresponds to half the height of a cylinder with the same volume as a unit sphere, and accounts, in approximation, for all possible paths of the e$^-$.

Considering the complexity of the photon absorption processes, the conversion into electrons, and how these electrons loose their kinetic energy, it is clear that such a definition is oversimplified.
However, it is sufficient to take into account the experimental results and provide a reasonable analysis.

We calculated the intensity of the current $i(x,z)$ taking into account the 2D Gaussian beam shape and
the linear absorption length on W for photons of 10 keV ($\mu$ = 5.8 $\mu$m).
The current is then given by:

\begin{equation}
	i(x,z)=\int{ I(x,z) \frac{1}{3} f(x,z) (1-e^{-t(x,z)/\mu_l}) dx} dz
	\label{eq:current}
\end{equation}

$I(x,z)$ is the Gaussian distribution of the photons, $f(x,z)$ takes into account only the number of photons that are absorbed at a distance smaller than $d_{eff}$ from the surface and \(t(x,z)\) is the length along which the absorbed photons produce measurable current.
Both $f(x,z)$ and $t(x,z)$  depend on the geometry of the tip as well as on the geometry of the incidence and therefore both depend on the parameter $d_{eff}$.
If the radius of the tip is smaller than $d_{eff}$ then $f=1$ and $t(x,z)$ is simply the thickness \(t = \sqrt{R(z)^2-x^2}\) with $R(z)$ being the radius of the tip at position $z$.
If the radius of the tip is much larger than $d_{eff}$ and the beam is incident on the tip perpendicular to it, then $t(x,z)= d_{eff}$.

On the basis of this calculation we can now estimate $d_{eff}$.
Figure \ref{tip_mesh}a, \ref{tip_mesh}b and \ref{tip_mesh}c show the agreement between measured and calculated current versus the tip position relative to the beam.
The only adjustable parameter is $d_{eff}$.
All other quantities are derived from experimental conditions.
From our experiments, we systematically obtain $d_{eff}$ of about 30 nm or less.
This is an important figure since it indicates how deep probe the surface is probed and gives an insight into the best possible lateral resolution.
Neglecting any issue of convolution with the tip shape, the best possible resolution in TEY collection cannot be better than this value, that in turn is material dependent.

As a final remark, we would like to point out that additional information is coming from the use of a lock-in in detection of the TEY signal.
In figure \ref{tip_mesh}d as we scan the $L_3$ edge of tungsten, we see the edge jump both in the  X and Y lock-in signals.
X is in phase with the signal, while Y is at $\pi/2$ with a value close to zero along all the scan except at the white line (the peak).
A tentative explanation might be that this is caused by the low energy electrons that cannot penetrate the barrier.

\section{XAS and AFM}

We discuss here the possibility of selecting an individual particle or a specific region of the sample and measure XAS (X-ray absorption spectroscopy) signal with the AFM tip.

\begin{figure}[h]
	\begin{center}
		\includegraphics[width=1\linewidth]{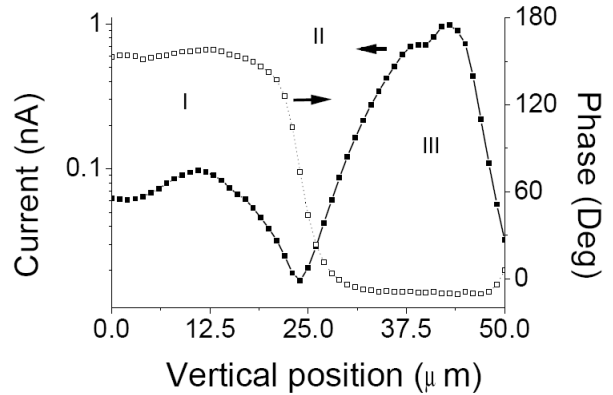}
		\caption{Evolution of the current as the beam is moved form the tip to the sample: (I) beam incident on the W tip, (II) beam passing the tip and the sample,  (III) beam illuminating sample. \vspace{-16pt}}
		\label{ZScan}
	\end{center}
\end{figure}

In figure \ref{ZScan} we illustrate the evolution of the detected current as the beam is moved form the tip into the sample.

At the beginning, we measure a signal due to beam on tip (region I). Corresponding phase is close to $180^o$.
As we further move the beam toward the sample, we observe a minimum corresponding to the position where the beam is passing between the tip and the sample (region II).
Moving further,
we end up illuminating the region of the sample below the tip resulting in a maximum of signal (region III) from the illuminated sample, the corresponding phase has shifted $\pi$ indicating a reversal in the current direction, which corresponds to detection of photoelectrons emitted by sample.

\begin{figure}[h]
	\begin{center}
		\includegraphics[width=1\linewidth]{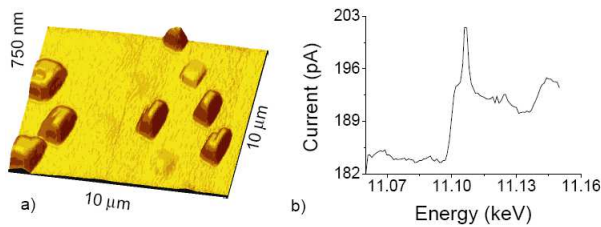}
		\caption{a) In-situ AFM topography image, b) Ge $L_3$ edge.} 
		\vspace{-16pt}
		\label{AFM_EXAFS}
	\end{center}
\end{figure}

After careful alignment of beam, tip apex and sample, we recorded simultaneously the topographic signal and the TEY, looking for some sort of chemical contrast when passing over one island.
We could not detect such a contrast since allthe emitted electrons within the footprint of the beam (fig. \ref{AFM_EXAFS} islands) are collected by the bare tungsten tip. As we will say below more effort has to be put in engineering tips for local measurements.

However, the EXAFS signal from the Ge islands was detected.
For this the tip was lifted of about 10$\mu$m, in order not to worry about AFM control.
The beam energy was varied across the Ge K edge and the current variation measured.
The result is presented in figure \ref{AFM_EXAFS}b where we plot the current flowing to the tip as a function of the energy.

From the data, it is possible to observe that the Ge islands were quite oxidized as indicated by the presence of a shoulder before the white line \cite{Greegor}.

From figure \ref{AFM_EXAFS}b it is observed that the jump is about 10 pA which corresponds to only 5\% of the signal.
Since $\approx$ 10 islands were illuminated, each one contributed about 1 pA to the edge.
Given the geometry of the islands and their distance to the tip, the  tip was collecting in a solid angle of $\approx$ $\pi/3$ Sr. Therefore only 1/10 of the total emission was measured.

With "smarter tips", (an example of which would be an STM tip isolated down to the apex with only a very small opening ($\approx 50 nm$) at the extremity), and being close to the particles (larger solid angle) we expect to detect islands at least 10 times smaller since signals on pA range and smaller are nowadays easily measured.
Moreover, from figure \ref{AFM_EXAFS}b, we see that most of the signal comes from the background.  
A "smarter tip" tip would dramatically reduce the background allowing a more effective amplification of the signal without saturation.

\section{Diffraction and AFM}

We discuss here how Bragg reflections can be measured through collection of diffracted beams by the tip.

\begin{figure}[h]
	\begin{center}
		\includegraphics[width=1\linewidth]{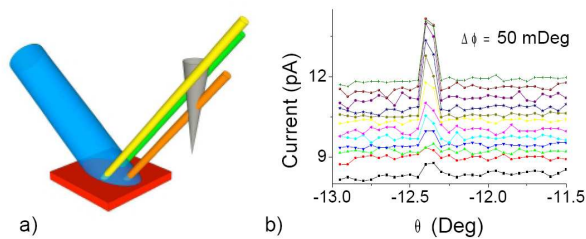}
		\caption{a) scheme illustrating how different areas can be resolved b) photocurrent from the W tip as a function of the incidence angle $\theta$ , and azimuthal $\phi$. \vspace{-16pt}}
		\label{diffraction}
	\end{center}
\end{figure}

In typical diffraction experiments, the detector is located about one meter away from the sample and collects photons in a very small solid angle, to provide a very high angular resolution.
The use of STM tips to detect the diffracted photons cannot give the same angular resolution, but instead, can give a very high \textit{spatial} resolution capable of resolving beams diffracted from different nanoareas as illustrated in figure \ref{diffraction}a.

The coarse alignment of the Ge (311) reflection for an energy of about 11keV corresponds to a $\theta$ angle of about $12^o$.
The fine alignment has been carried out using exclusively the AFM tip scanning $\theta$ for a set of different azimuthal angles $\phi$ (fig. \ref{diffraction}b).

\begin{figure}
	\begin{center}
		\includegraphics[width=1\linewidth]{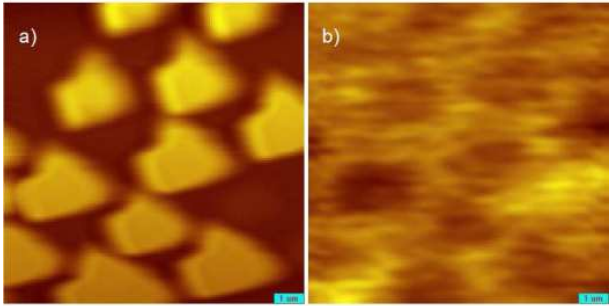}
		\caption{ (a) AFM image; (b) current image: the contrast in the current image in which we see the Ge islands as holes, is because the tip collects the diffracted x-rays which subtracts from the photoelectrons. \vspace{-16pt}}
		\label{Ge_Bragg}
	\end{center}
\end{figure}

The Bragg conditions are meet at $\theta = 12.35$ Deg and $\phi = 88.7 Deg$.
As an example, the results obtained using the tip as a photon detector while acquiring simultaneously an AFM topography image are here reported.
When in Bragg conditions, the current signal shows a significant contrast (fig.\ref{Ge_Bragg}b) that disappears completely when the sample is not in Bragg conditions.

Unlike the absorption case a small contrast could be obtained.
The reason is because the photons are emitted in a very well defined direction.
From the contrast in figure \ref{Ge_Bragg}b,
corresponding to 0.2 \% ($\approx 2pA$) of the total signal,
we can identify each of the islands shown in the topographic image of fig. \ref{Ge_Bragg}a.

The particles are seen as darker holes, since the diffracted photons impinging on the W tip extract electrons away, which causes a decrease in the overall current which is mainly due to the electrons photoemitted from the sample.

The AFM image is distorted because of a tip effect.
The sample being the same as in figure \ref{AFM_EXAFS}a,
the systematic deformation has to be attributed to the tip which became blunt after several hours of usage.

Here again, the contrast observed as the tip cuts through the array of diffracting beams would be strongly enhanced using a "smarter tip" since it would strongly diminish the more or less constant background of photoelectrons.
It would also prevent photoelectrons generated in the tip, far from its apex, to escape away increasing the angular resolution in the scattering plane.

\section{Perspectives}

In a different experiment with other groups, we are also trying to measure diffraction from an individual nano crystal while the AFM tip mechanically interacts with it. 

Finally, with the current trend toward SR "nanobeams" it will be harder and harder to tell which part of the sample is being investigated.
It is our believe that the techniques described here will have an important impact in this context.

\section{Acknowledgments}

We thank D. Wermeille from ID3 ESRF for his help at the beam line.
We thank T. H. Metzger, C. Mocuta and K. Mundboth from ID1 at ESRF for discussions.
This work was supported by the E.U. FP6 program, under contract STRP 505634-1 X-tip, and the Portuguese Foundation for Science and Technology for its financial support.

\end{document}